\documentclass[twocolumn,english,aps,pre,superscriptaddress,showpacs]{revtex4}
\usepackage[T1]{fontenc}
\usepackage[latin9]{inputenc}
\usepackage{amsmath}
\usepackage{amssymb}
\usepackage{pict2e}
\usepackage{esint}
\usepackage{graphicx}
\makeatletter
\@ifundefined{textcolor}{}
{%
 \definecolor{BLACK}{gray}{0}
 \definecolor{WHITE}{gray}{1}
 \definecolor{RED}{rgb}{1,0,0}
 \definecolor{GREEN}{rgb}{0,1,0}
 \definecolor{BLUE}{rgb}{0,0,1}
 \definecolor{CYAN}{cmyk}{1,0,0,0}
 \definecolor{MAGENTA}{cmyk}{0,1,0,0}
 \definecolor{YELLOW}{cmyk}{0,0,1,0}
 }

\usepackage{nicefrac}\usepackage{dsfont}
\usepackage[normalem]{ulem}\@ifundefined{definecolor}
 {\usepackage{color}}{}

\makeatother

\usepackage{babel}
\begin{document}

\title{Active biopolymer networks generate scale-free but euclidean clusters}

\author{M. Sheinman\textsuperscript{1,2}, A. Sharma\textsuperscript{1}, J. Alvarado\textsuperscript{3,4}, G. H. Koenderink\textsuperscript{3}, F. C. MacKintosh}
\address{\textsuperscript{}Department of Physics and Astronomy, VU University, Amsterdam, The Netherlands\\
\textsuperscript{2}Max Planck Institute for Molecular Genetics, 14195 Berlin, Germany\\
\textsuperscript{3}FOM Institute AMOLF, Science Park 104, 1098 XG Amsterdam, The Netherlands\\
\textsuperscript{4}Department of Mechanical Engineering, Hatsopoulos Microfluids Laboratory, Massachusetts Institute of Technology, Cambridge, Massachusetts 02139, United States}

\date{\today}
\begin{abstract}
We report analytical and numerical modelling of active elastic networks, motivated by experiments on crosslinked actin networks contracted by myosin motors. Within a broad range of parameters, the motor-driven collapse of active elastic networks leads to a critical state. We show that this state is qualitatively different from that of the random percolation model. Intriguingly, it possesses both euclidean and scale-free structure with Fisher exponent smaller than $2$. Remarkably, an indistinguishable Fisher exponent and the same euclidean structure is obtained at the critical point of the random percolation model after absorbing all enclaves into their surrounding clusters. We propose that in the experiment the enclaves are absorbed due to steric interactions of network elements. We model the network collapse, taking into account the steric interactions. The model shows how the system robustly drives itself towards the critical point of the random percolation model with absorbed enclaves, in agreement with the experiment.
\end{abstract}
\maketitle
\section{Introduction}
Motivated by living matter, internally driven elastic networks have been intensively studied during the last decade. One example of such an active network is the actin cortex of living cells, which is composed of cytoskeletal actin filaments, crosslinks, and myosin motors. Taken together, these cellular proteins form an elastic network with contractility induced by non-equilibrium motor stresses \cite{brangwynne2008cytoplasmic,levayer2012biomechanical}. The motors generate forces which propagate through the network and govern viscoelastic properties, structure, shape and movement of cells \cite{humphrey2002active,mizuno2007nonequilibrium,mackintosh2008nonequilibrium,koenderink2009active,liverpool2009mechanical,schaller2010polar,broedersz2011molecular,wang2012active,bertrand2012active,sheinman2012actively}. 

Active elastic networks, in general, and the cytoskeleton, in particular, are still poorly understood. Studying active remodeling of networks has been particularly challenging. Recent experiments on reconstituted acto-myosin networks \textit{in vitro} have demonstrated restructuring of the network by the motors. Motor activity can lead to network failure and its subsequent collapse to isolated clusters \cite{kohler2011structure,Silva2011active,murrell2012f}. In certain cases the collapse of the network robustly leads to scale-free structures, as reported in Ref. \cite{alvarado2013myosin}. In the same study modelling of the network collapse traced the robustness of the resulting structure to a motor-driven reduction of network connectivity. In the current study we focus on the properties of the scale-free structures and explain how one can generate, seemingly, paradoxical scale-free and, yet, compact clusters. We build a simple model, which allows us to analyze the influence of steric interactions within the network on the resulting state. This model with steric interactions results in a robust collapse to a state, qualitatively similar to the experimental one. 

\section{Experimental results}
To gain insight into the physical properties of the active elastic networks, we focus on the properties of the clusters' structure and statistics, using a combination of experimental, analytical and numerical approaches. Experimentally, we study the motor-driven collapse of model cytoskeletal system, composed of actin filaments, fascin cross-links and myosin motors, using the approach, developed in Ref. \cite{alvarado2013myosin}.
In short, cluster sizes were measured by recording images of contracting actomyosin networks and tracking cluster expansion in reverse time using a customized image analysis algorithm. Cluster size distributions were determined for three sample regimes (controlled by crosslink density): many small clusters, clusters with scale-free distributed sizes, and few large clusters. 

We find that the second regime surprisingly spans a wide range of experimental parameters and is manifested in a scale-free cluster size distribution:
\begin{equation}
n_s \sim s^{-\tau},
\label{ns1}
\end{equation}
as has been observed experimentally in Ref. \cite{alvarado2013myosin}.
Here $n_s$ is the number of clusters with initial area (for a compact cluster its area is proportional to its mass) $s$ and $\tau$ is the Fisher power-law exponent.
This suggests a connectivity percolation phenomenon, where the clusters are also distributed in a scale-free manner at the percolation point. However, the details of the experimental results cannot be explained by a \textit{random} percolation model.

\begin{figure}[htb]
\includegraphics[width=0.5\columnwidth]{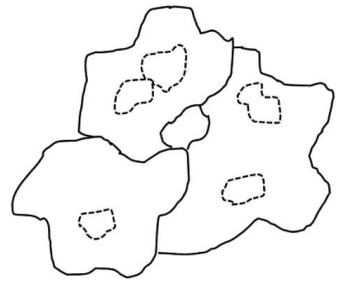}
\caption{Illustration of enclaves---clusters fully surrounded by other clusters. In this example the enclaves are marked by dashed boundaries and non enclaves by solid boundaries.}
\label{Enclaves}
\end{figure}

\begin{figure}[htb]
\centering

  \begin{tabular}{@{}cccc@{}}
  \includegraphics[width=.23\textwidth]{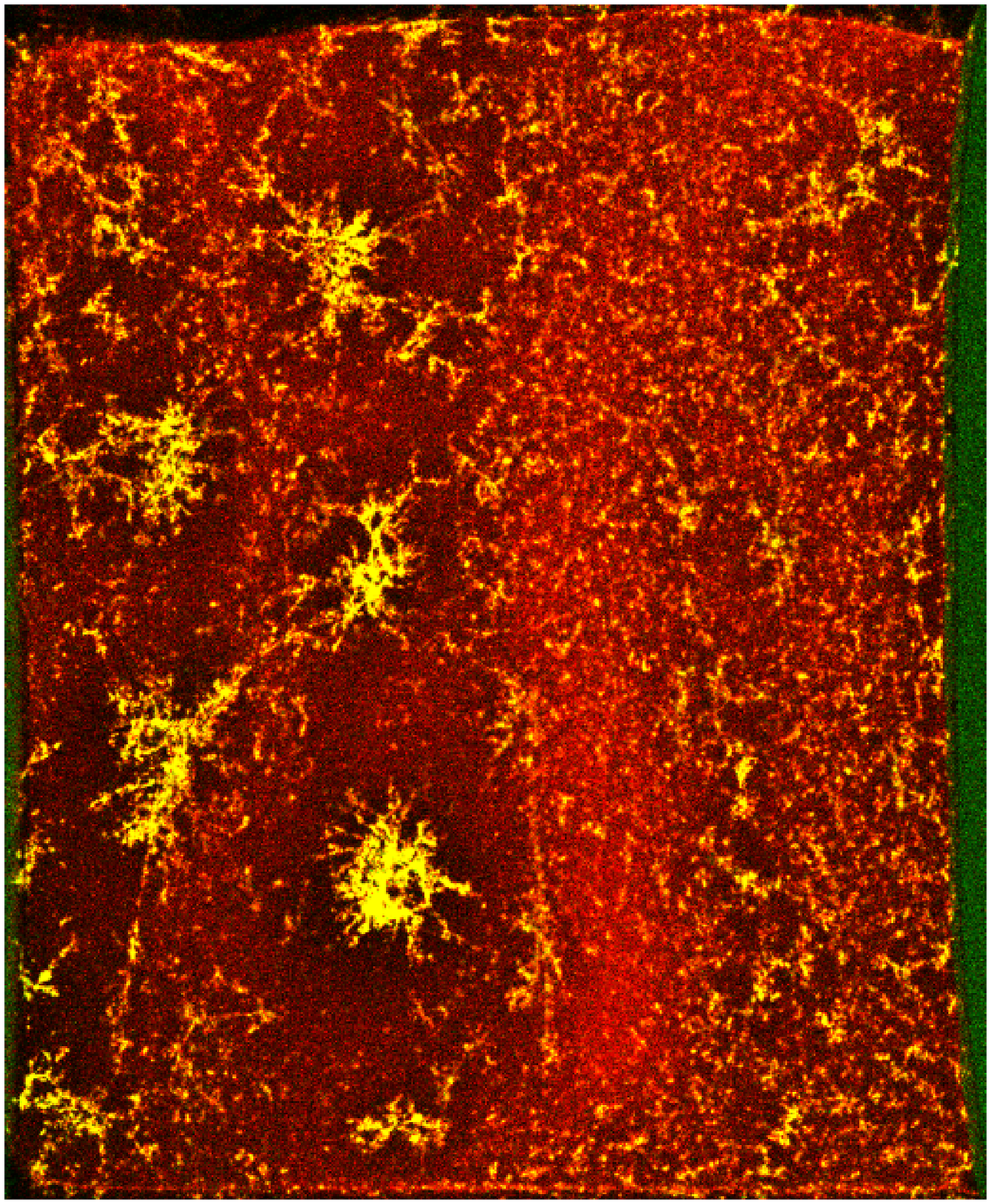}
    \put(-120,136){\fcolorbox{white}{white}{(a)}} &
    \includegraphics[width=.23\textwidth]{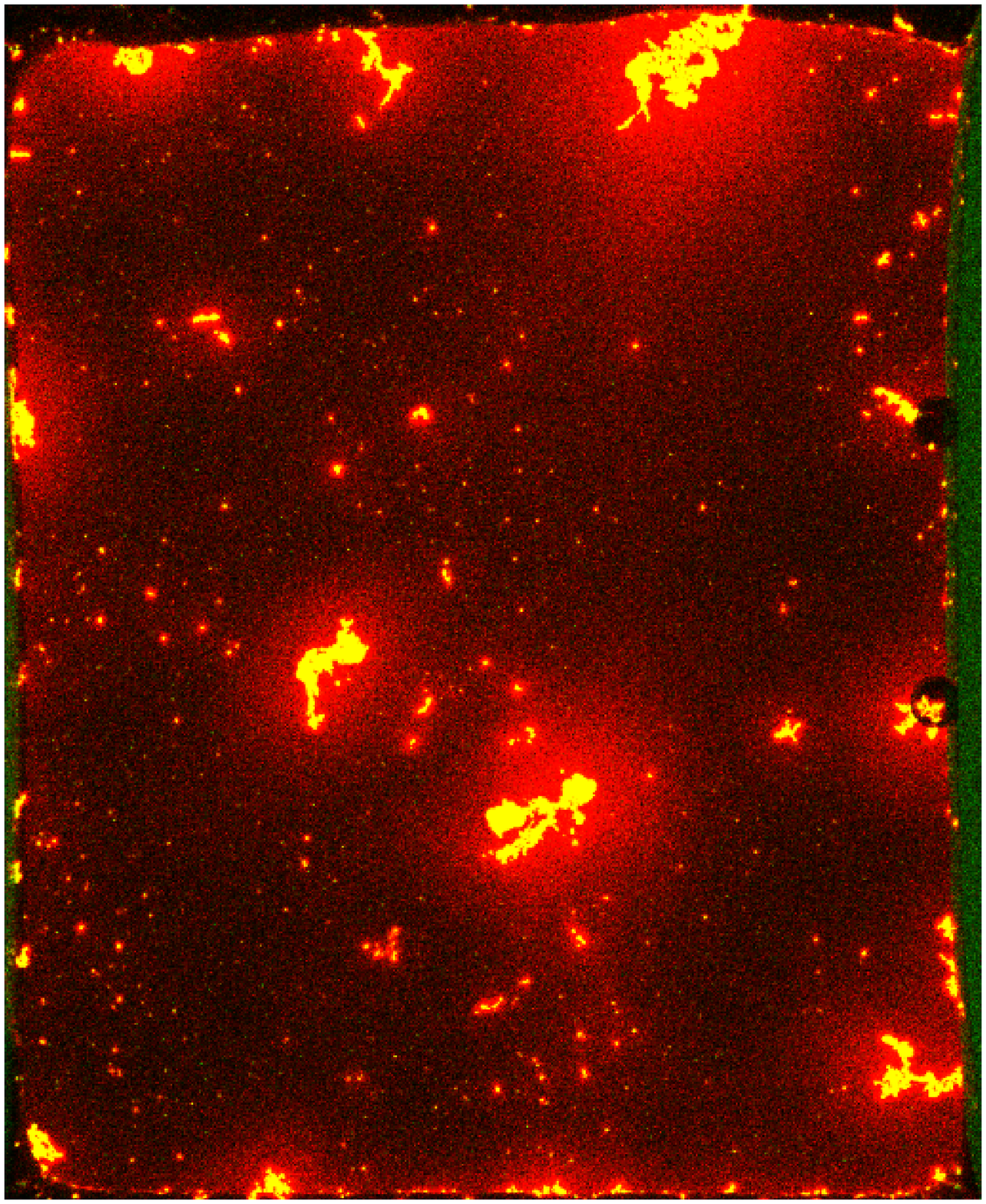} 
        \put(-120,135){\fcolorbox{white}{white}{(b)}}& \\
    \includegraphics[width=.23\textwidth]{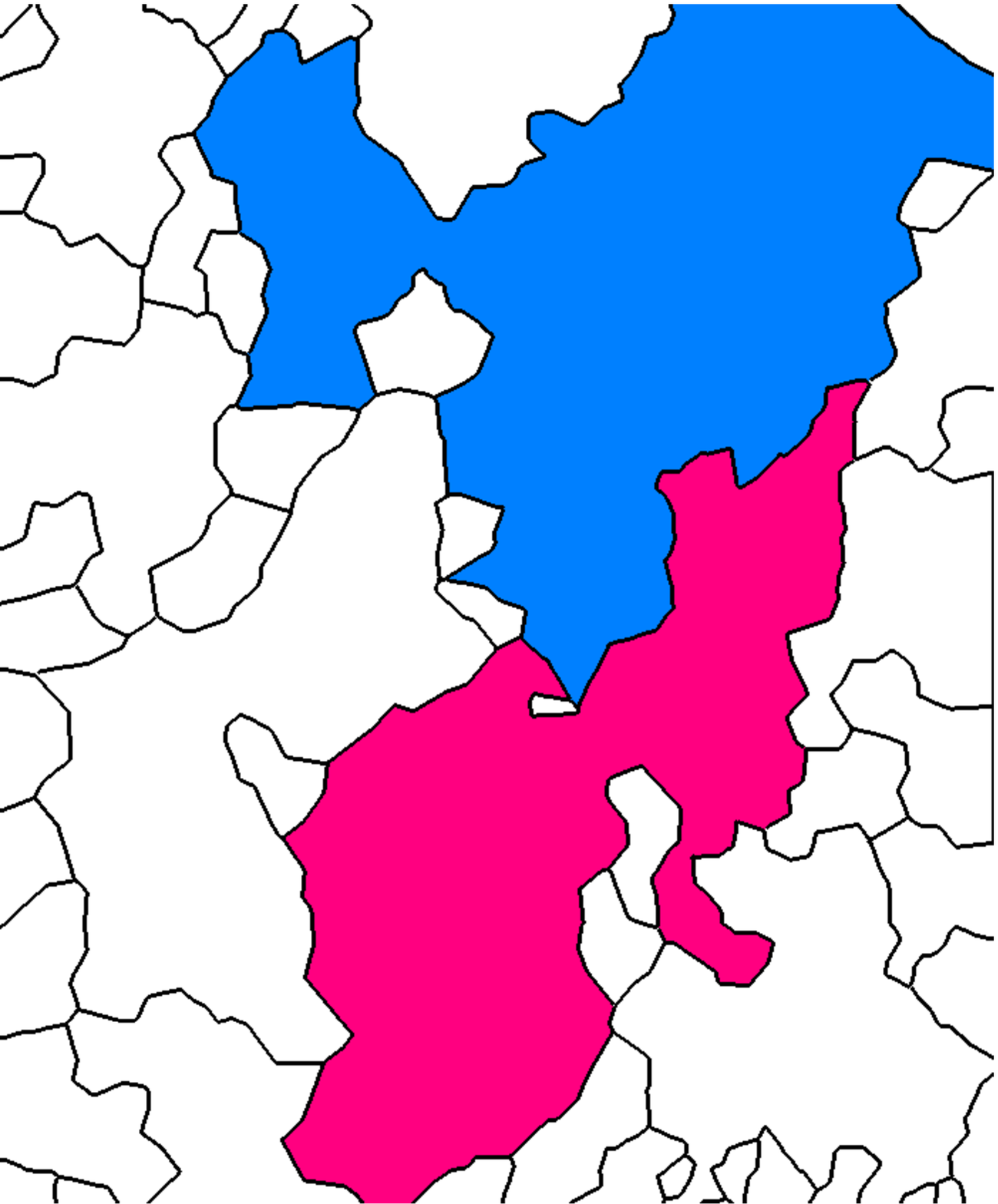} 
    \put(-125,135){\fcolorbox{white}{white}{(c)}}&
    \includegraphics[width=.23\textwidth]{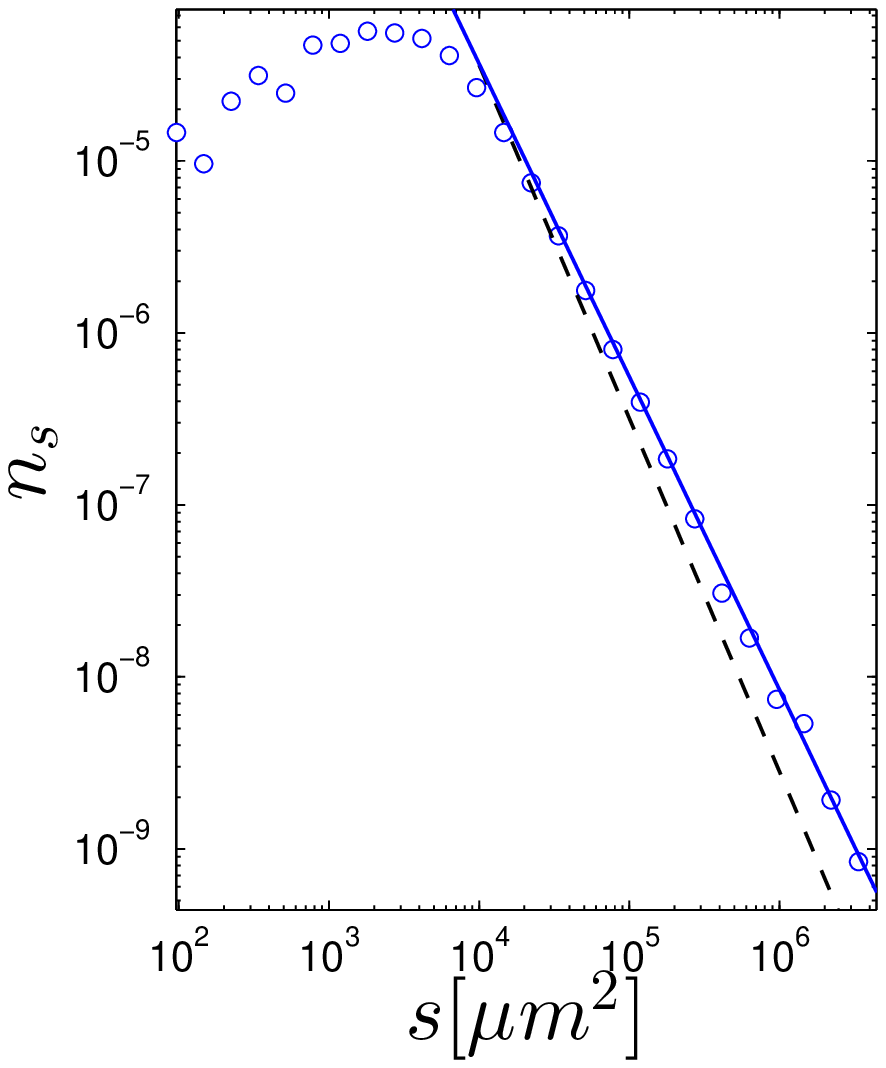} 
    \put(-96,132){(d)}
  \end{tabular}
\caption{Experimental results of a fascin-crosslinked actin network, collapsed by myosin motors. The concentrations of crosslinks, actin and molecular motors are given by $0.24$, $12$ and $0.12$ $\mu M$, respectively. The dimensions are $3.18mm \times 2mm$, while the sample thickness is $80 \mu m$. (a) The network after $2$ min. (b) The network after $104$ min. The movie of the collapse can be seen in Sup. Mat. (c) Initial configuration of the collapsed clusters. Colours indicate the largest (blue) and the second-largest (pink) clusters. (d) Histogram (circles) of cluster initial areas, averaged over 26 samples. For the critically connected regime, the data is statistically more consistent ($1.4$ standard errors from the Hill estimator of $\tau=1.91 \pm 0.06$ \cite{clauset2009power}) with a power-law distribution with a NEP model's Fisher exponent from Eq. \eqref{tau} (solid line). The agreement of the data with the random percolation model Fisher exponent $\tau=187/91$, indicated by the dashed line, is significantly worse ($2.4$ standard errors from the Hill estimator).}
\label{Jose}
\end{figure}

\begin{figure}[htb]
\centering
  \begin{tabular}{@{}cccc@{}}
    \includegraphics[width=.23\textwidth]{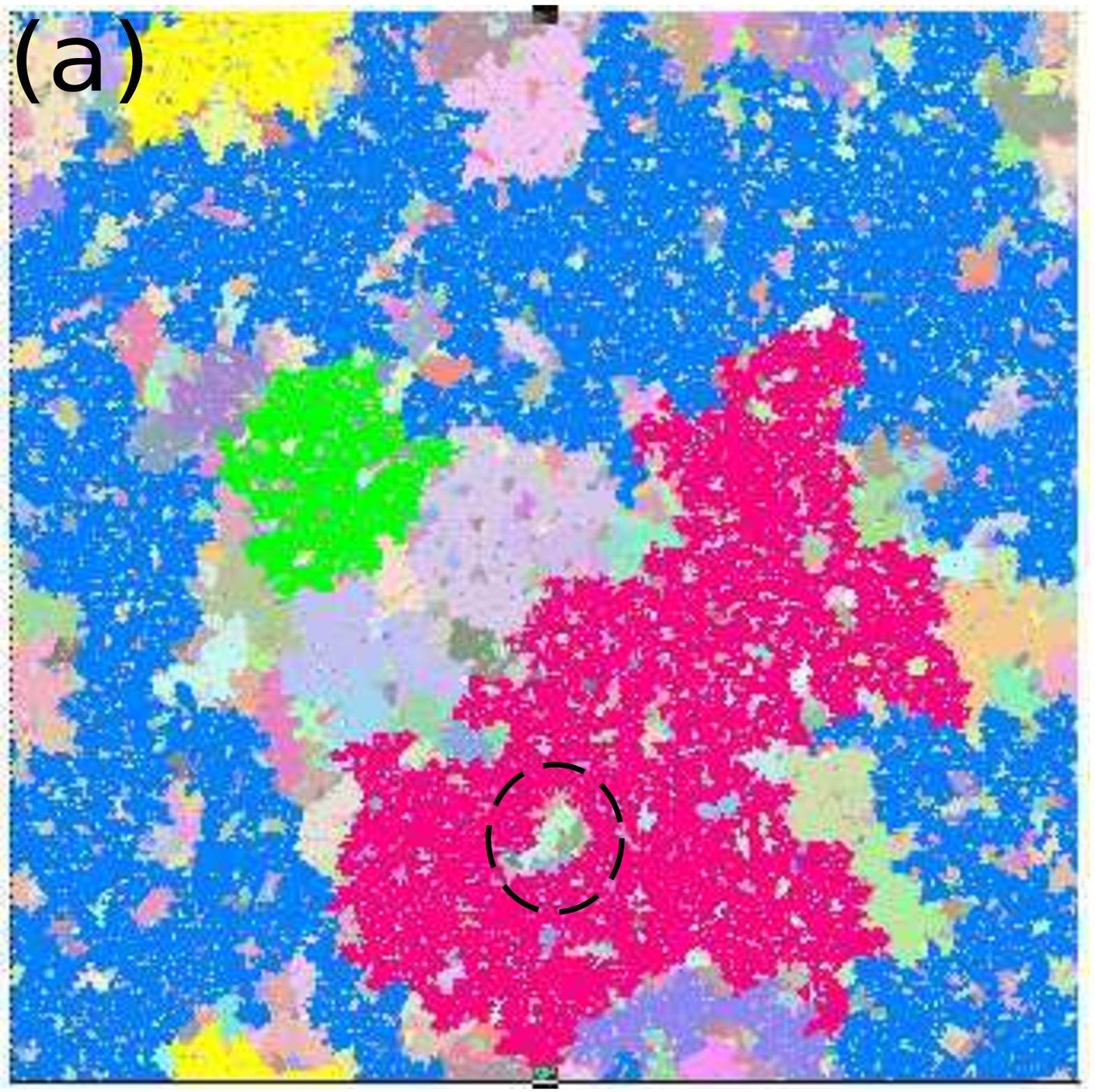} &
    \includegraphics[width=.23\textwidth]{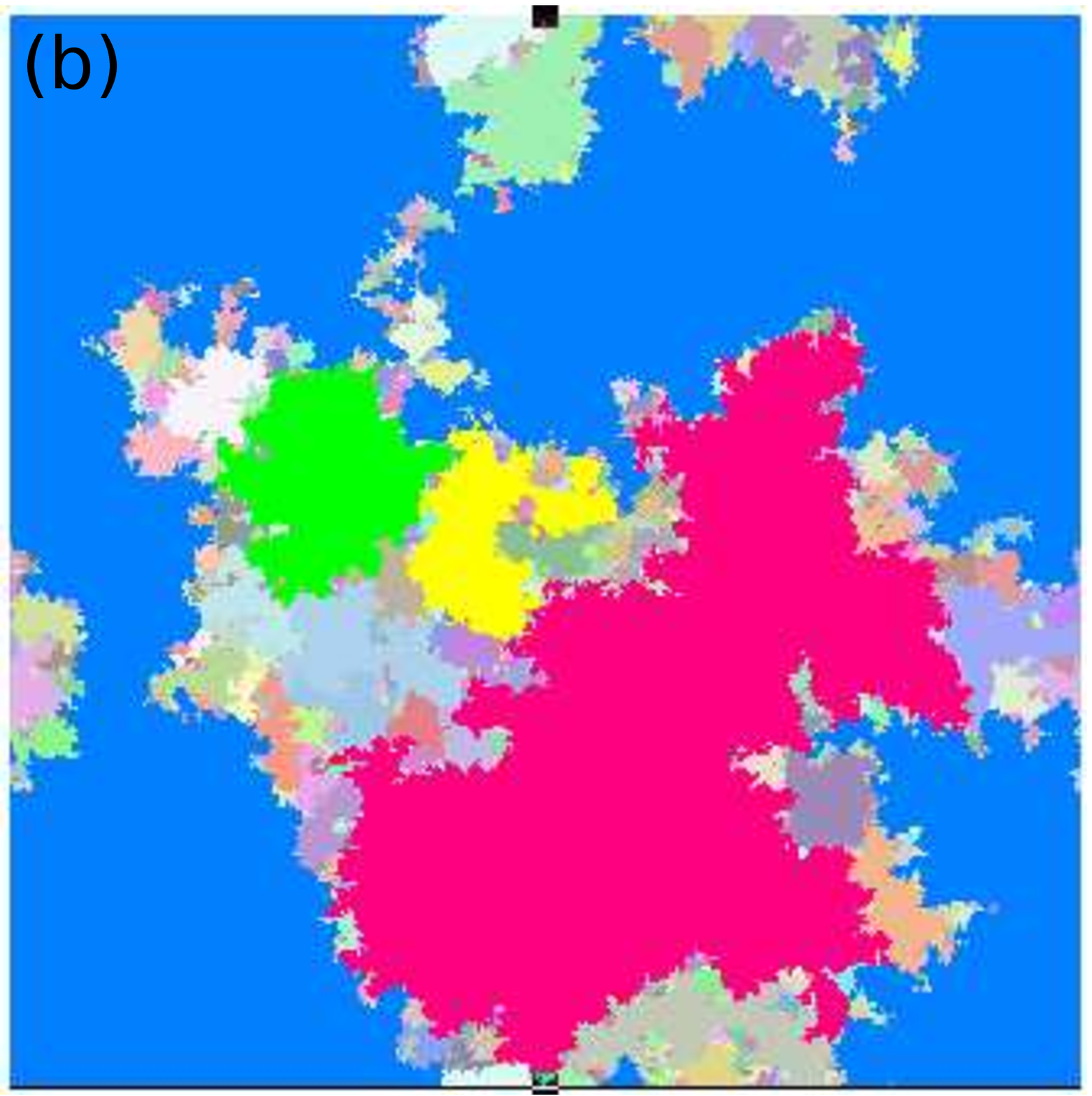} & \\
    \includegraphics[width=.23\textwidth]{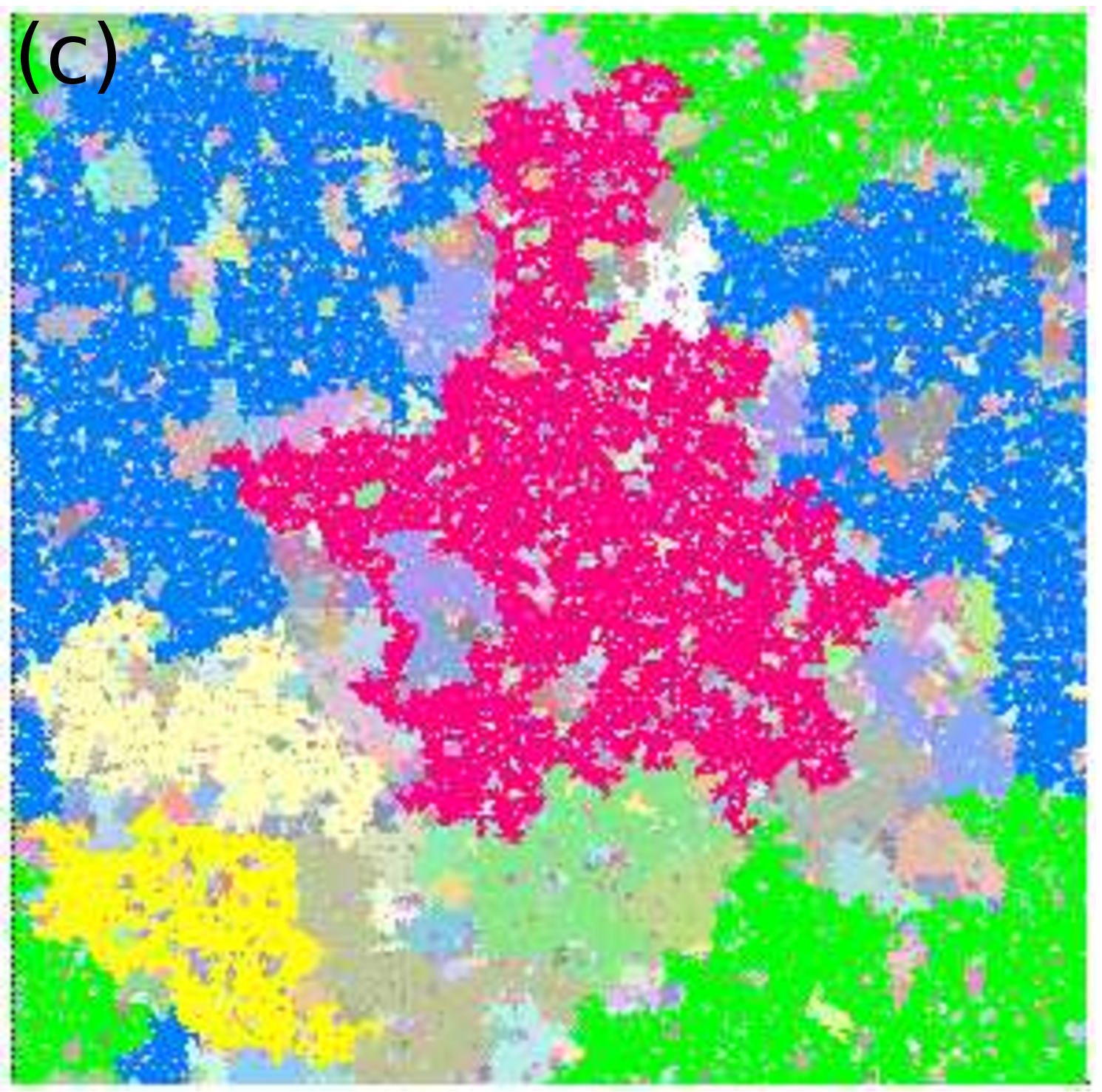} &
    \includegraphics[width=.23\textwidth]{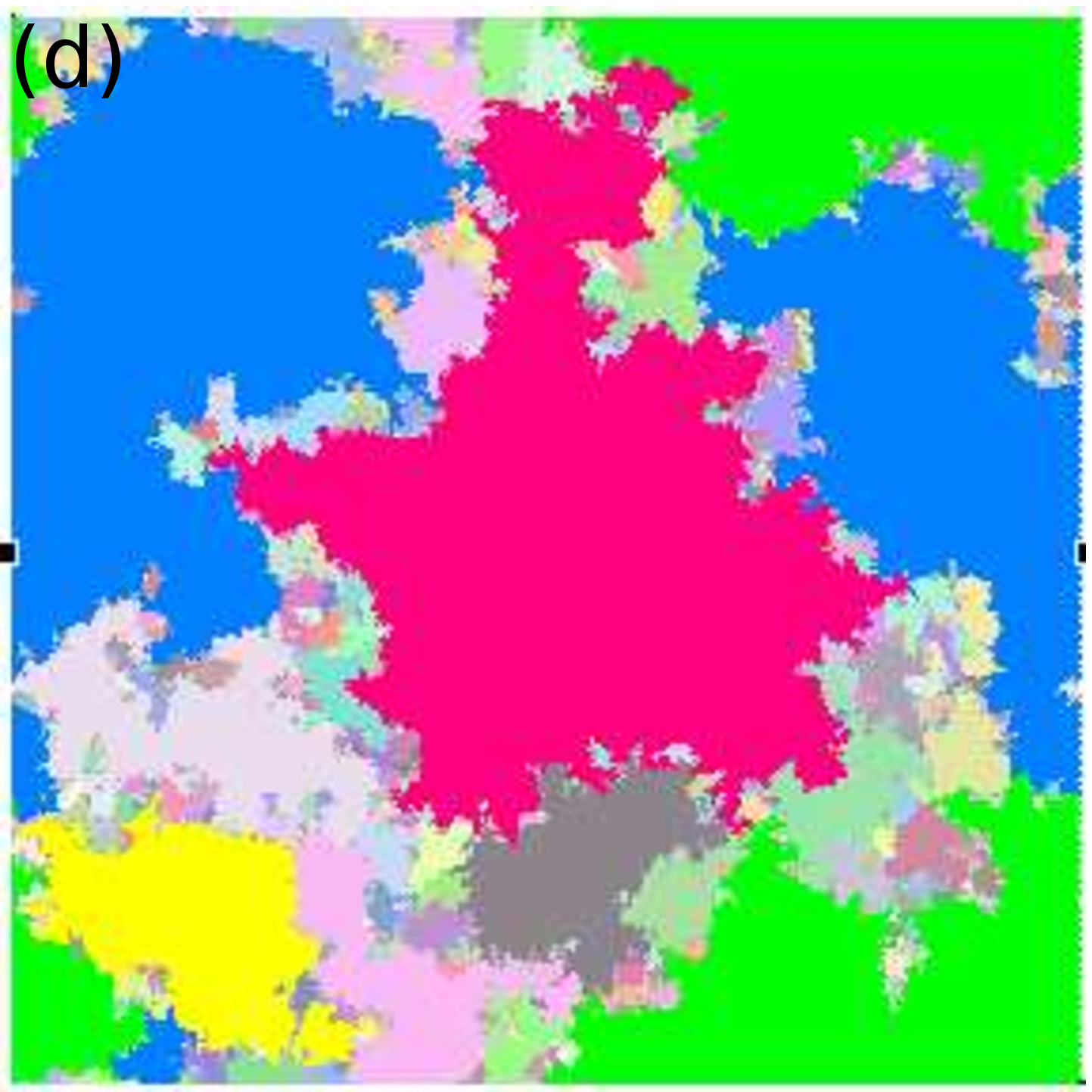} & \\
  \end{tabular}
\caption{(a,b) Clusters' structure  at the critical point of the random percolation model without (a) and with (b) absorbed enclaves---NEP model. Dashed circle in (a) indicates an enclave, absorbed in (b). (c,d) Modelling of the motor-driven collapse with and without steric interactions. Initial configuration of collapsed clusters modelled without (c) and with (d) steric interactions. Each cluster is indicated by a different grey level (color in the online version) on each plot.}
\label{InitialAreas}
\end{figure}

Firstly, as indicated in Ref. \cite{alvarado2013myosin}, the robustness of the scale-free collapsed structure does not exist in a random percolation model, where the criticality is apparent only in a very narrow vicinity near the percolation point. 
Secondly, the properties of the scale-free structure in the experiment and in a random percolation model are qualitatively different. In the random percolation model, at the percolation transition, clusters possess many scale-free enclaves---clusters fully surrounded by another cluster (see Fig. \ref{Enclaves}). The presence of enclaves makes the clusters non-compact. In fact, the clusters are so porous, that their mass scales sublinearly with their volume. Such structures are random fractals, with the fractal dimension $d_\mathrm{f}'=91/48<2$. Since the density of the largest cluster vanishes in the thermodynamic limit, the percolation transition  is continuous \cite{stauffer1994introduction,sheinman2014Discontinuous}. An example of a network configuration at the critical point of the random percolation model with such fractal clusters can be seen in Fig. \ref{InitialAreas}(a). 
However, the properties of the scale-free structure in the experiment are different from those at the percolation transition point of the random percolation model. In the experiment the motors collapse the network to multiple small clusters (see Fig. \ref{Jose}(a,b) and movie in the Sup. Mat.). The initial configuration of the collapsed clusters possesses no enclaves, as shown in Fig. \ref{Jose}(c). The absence of enclaves in the experiment suggests non-fractal, compact, euclidean clusters. 
Moreover, the Fisher exponent, $\tau$, of the clusters' initial area distribution in Eq. \eqref{ns1} in the experiment is smaller than $2$ (see Fig. \ref{Jose}(d)). This is inconsistent with a random percolation model. There the order parameter is continuous at the percolation transition, implying $\tau'=187/91>2$ \cite{stauffer1994introduction}. In fact, the Fisher exponent and fractal dimension are related via a hyperscaling relation \cite{stauffer1994introduction}. Therefore, it is expected that violation of $d_\mathrm{f}<2$ (as shown for the experimental system in Fig. \ref{Jose}(c)) also implies violation of $\tau>2$ (as shown for the experimental system in Fig. \ref{Jose}(d)). 

To get a structure with such properties, one needs a seemingly paradoxical percolation model with a discontinuous, yet critical transition \cite{sheinman2014Discontinuous}. As shown in Ref. \cite{sheinman2014Discontinuous}, the modified percolation model with absorbed enclaves possesses such properties. Here we demonstrate that this no-enclaves percolation (NEP) model naturally emerges by taking into account steric interactions within the elastic network during the collapse process. We present a physical mechanism that robustly drives the system towards the critical point of the NEP model, with euclidean clusters and, yet, critical, scale-free cluster size distribution, similarly to the experiment. We begin with a description of the NEP model and review of its properties, which are relevant for the experiment.

\section{No-enclaves percolation (NEP) model}
During the collapse process of active networks, due to excluded volume interactions, the enclaves of a cluster collapse to the same point as their surrounding cluster (see Fig. \ref{EnclaveInclusion} for illustration). In the experiment the initial configurations of the clusters were reconstructed, starting from the collapsed state. The collapsed state of a large cluster contains all its enclaves (see Fig. \ref{EnclaveInclusion} for illustration). This makes the reconstructed initial configuration of all clusters enclave-free (see Fig. \ref{Jose}(c)). This suggests, that the no-enclaves percolation (NEP) model, which has been introduced and analysed in Ref. \cite{sheinman2014Discontinuous}, can explain the universal properties of the experiment. The NEP model is the random percolation model, in which all the enclaves are absorbed in their surrounding clusters (see Fig. \ref{InitialAreas}(a,b)). We turn now to a description of this model and its relevant properties.

Consider a $2$-d lattice of massive nodes. Each pair of nearest-neighbouring nodes can be connected by a massless bond. Connectivity---the fraction of occupied bonds---is denoted by $p$.
The absorption of enclaves in their surrounding clusters makes the structure of the resulting, compact clusters euclidean, with fractal dimension $ d_\mathrm{f}=d=2 $. 
In contrast to the random percolation model, where $d_\mathrm{f}=91/48<d$ and the mass of the largest cluster at the percolation transition, $p=p_c$, scales as $M^{d_\mathrm{f}/d}$, in the NEP model the mass of the euclidean largest cluster scales linearly with the system size, $M$. Since the strength (mass fraction) of the largest cluster, $P$, in the NEP model does not vanish in the thermodynamic limit, the percolation transition is discontinuous as a function of the connectivity. However, the cluster mass distribution is still scale-free, with a qualitatively different Fisher exponent from the one in the random percolation model.

It has been shown analytically, that the Fisher exponent of the NEP model is bounded from above by $2$ \cite{sheinman2014Discontinuous}. 
In contrast, in the random percolation model, the Fisher exponent has to be larger than $2$ because of the fractality of the clusters \cite{stauffer1994introduction}. The Fisher exponent of the NEP model violates this constraint, such that the clusters' mass distribution scales as
\begin{equation}
n_s \sim s^{-\tau}
\label{ns}
\end{equation}
with $\tau<2$.
This is due to the compactness of the enclave-free clusters and different properties of the percolation transition in this model.

\begin{figure}[tb]
\centering
\centering
  \begin{tabular}{@{}cccc@{}}
    \includegraphics[width=.23\textwidth]{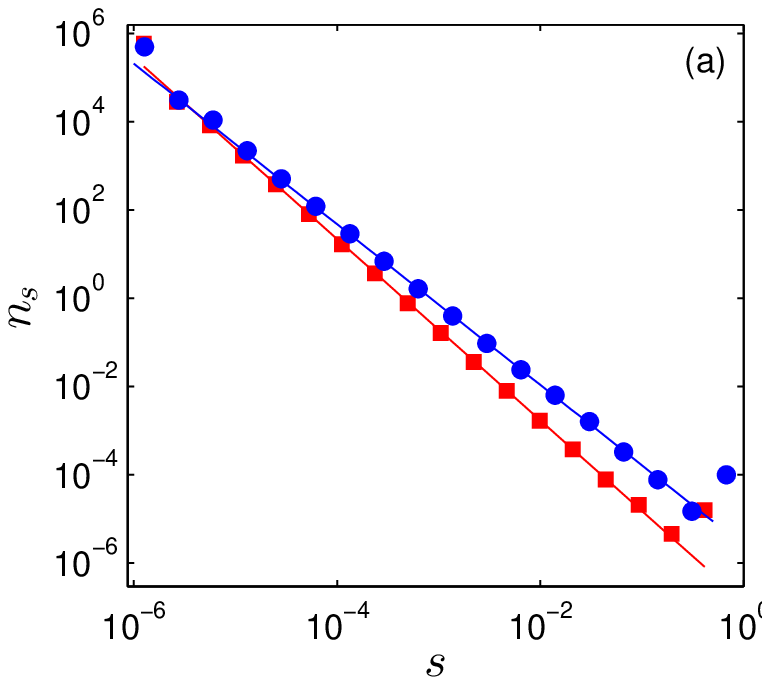} &
    \includegraphics[width=.23\textwidth]{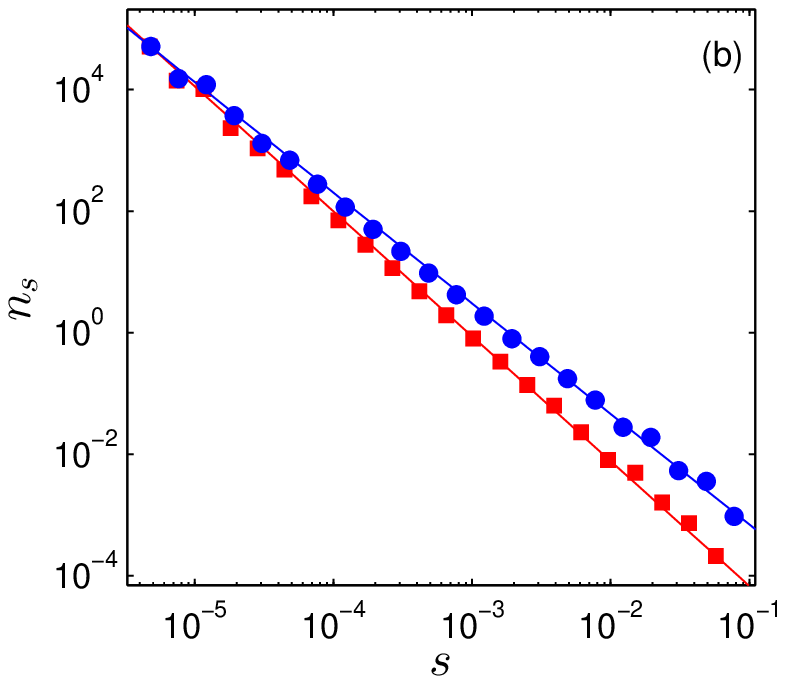} &
  \end{tabular}
\caption{Cluster mass (in units of a node's mass) distributions. (a) Critical point of the random percolation model without (squares) and with (circles) absorption of the enclaves into their surroundings (NEP model). (b) Collapsed state of the collapse model without (squares) and with (circles) steric interactions.  The fits to squares in both panels are with the power-law of the random percolation model  $ \tau=187/91>2 $ while for the circles the fit is with the NEP model's Fisher exponent from Eq. \eqref{tau}. }
\label{Distributions}
\end{figure}

\begin{figure}[tb]
\centering
\includegraphics[width= \columnwidth]{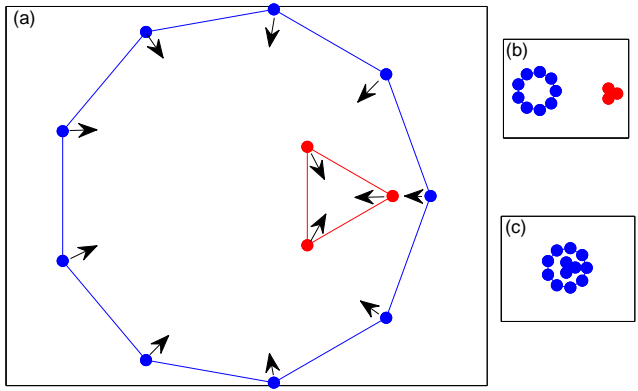}
\caption{An illustration of the influence of steric interactions on the collapsed state properties. The initial state presented in (a) without steric interactions collapses to two clusters with masses $ 3 $ and $ 9 $, as show in (b). Due to steric interactions the enclave is trapped in its surroundings, leading to a collapsed state with one cluster of mass $ 12 $, as shown in (c).}
\label{EnclaveInclusion}
\end{figure}

Numerically the Fisher exponent of the NEP model has been estimated in Ref. \cite{sheinman2014Discontinuous} to be
\begin{equation}
\tau=1.82 \pm 0.01,
\label{tau}
\end{equation}
as shown in Fig. \ref{Distributions}(a).
Remarkably, this value of $\tau$ is consistent with the value of the Fisher exponent, measured in experiments, as shown in Fig. \ref{Jose}(d).

In sum, the final state of the collapsed acto-myosin network possesses the universal properties (absence of enclaves and the value of the Fisher exponent) of the NEP model and not the random percolation model. In the following we present a minimalistic model of the collapse and show how the system can \textit{robustly} drive itself towards the critical point of the NEP model, in agreement with the experiment.

\section{The model of the collapse}
To model the failure and collapse of a network, induced by internal stresses, we use a spring network, initialized on a $ d=2 $-dimensional triangular lattice with periodic boundary conditions and total mass (or, equivalently, number of nodes) of $ L \times L=M$. Each pair of nearest-neighbouring nodes is initially connected by an elastic spring. 

Within our coarse-grained approach motor activity is reflected in two ways. 
Firstly, motors apply contractile forces on the network nodes \cite{huxley1969mechanism,mizuno2007nonequilibrium,koenderink2009active,dasanayake2011general,lenz2012contractile}. This we implement by setting the rest length of each spring to zero. Thus, the initial state, when springs' lengths are equal to the lattice constant, is under stress, such that a rupture of the network leads to its collapse. 

Secondly, motors reduce local connectivity by severing the filaments \cite{murrell2012f} or inducing unbinding of a crosslink \cite{ishikawa2003polarized}. Within our coarse-grained approach we implement this reduction of local connectivity by breaking the springs one by one in a random fashion. When a spring is broken the nodes, initially connected by the spring, disconnect. The mass of the spring is equally distributed between the nodes, such that the total mass of the network is conserved during the collapse process. 

\begin{figure}[tb]
\centering
\includegraphics[width= \columnwidth]{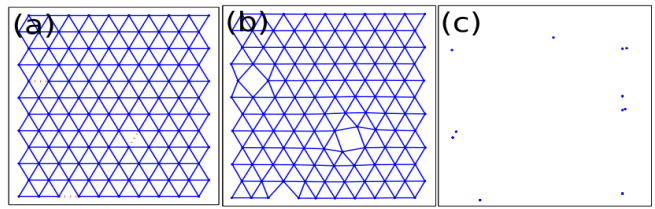}
\caption{An example of the collapse process of a small ($M=10 \times 12$) network. (a) initial state of the network. Dashed lines indicate first three springs that have been broken. (b) state of the network after first three breaks. (c) final state of the collapsed network. Note, that during the process the total mass (total number of nodes) is conserved. In the final state of the network all the mass is concentrated in 11 clusters.}
\label{Fig1}
\end{figure}

We quasistatically propagate the network, assuming that the breaks are sufficiently rare, such that the network balances all the elastic forces between any two subsequent breaks. An example of the process is show in Fig. \ref{Fig1} and the movie in Sup. Mat.

\begin{figure}[htb]
\centering
    \includegraphics[width=\columnwidth]{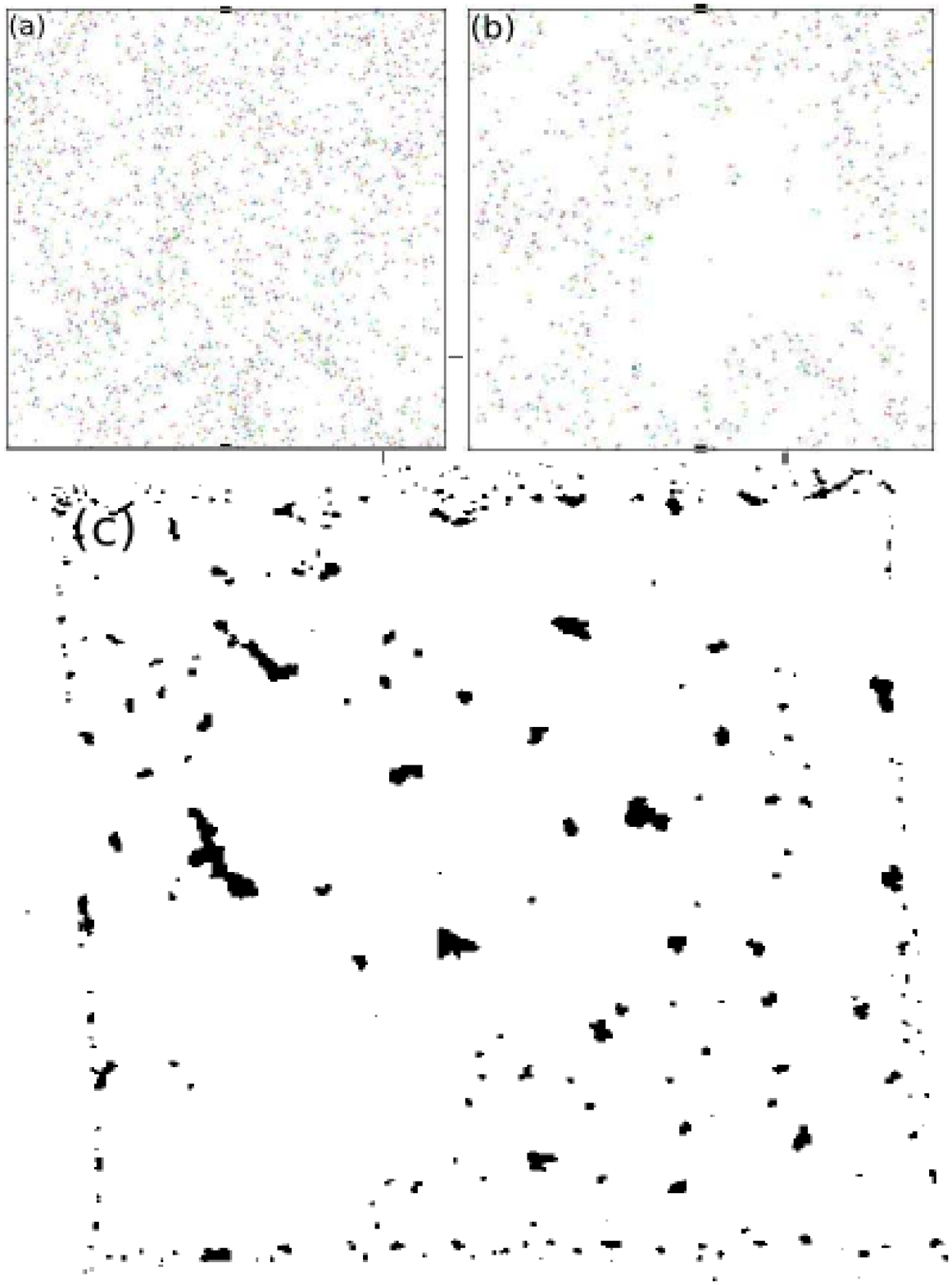} 
\caption{(a,b) Modelling of the collapse. The final state of ruptured collapsed network with $M=400 \times 400$ nodes, without (a) and with (b) steric interactions. Each point contains the nodes of a collapsed cluster. (c) The final state (converted to binary picture) of the collapse in the experiment (see caption of Fig. \ref{Jose} for experimental conditions). Note the large voids in the case of the modelling with steric interactions (b) and the experiment (c).}
\label{FinalState}
\end{figure}

The final state of the model's evolution is $ M $ disjointed nodes (see Figs. \ref{Fig1}(c) and \ref{FinalState}(a)) because all the springs are broken. However, many of the nodes, although not connected by springs, are located at the same \textit{geometrical} point in space. Similarly to the experiment, we define a collection of nodes in the same geometrical location in the final, collapsed state as a \textit{cluster}.

During the evolution of the network failure and collapse, the bond-breaking events in non-percolating subnetworks have no effect on the final state. This is because every finite subnetwork is collapsed to a point after it is detached from the infinite, percolating subnetwork, due to the quasistaticity assumption. Therefore, only the breaking events within the percolating subnetwork are relevant for the configuration of the clusters in the final state.

In addition to the elastic forces we also take into account steric interactions within the network, which prevent geometric overlap of different filaments in the network. In $2$-dimensional systems these interactions trap enclavic clusters in their surrounding clusters (see Fig. \ref{EnclaveInclusion} for illustration). As we show below, this qualitatively impacts the collapse process and its final state. To emphasize the importance of steric interactions, we first analyze the phantom version of our model of collapse ignoring them.

\section{Results}
\subsection{Phantom version of the collapse model}
Without steric interactions the mass, $ s$, of collapsed clusters (in units of nodes' mass) in the final state of the phantom model is found to be distributed in a scale-free fashion
\begin{equation}
n_s \sim s^{-\tau'}
\end{equation} 
with a Fisher exponent, consistent with the one of the random percolation model, $ \tau' = 187/91 $ \cite{den1979relation,nienhuis1982exact} (see Fig. \ref{Distributions}(a,b)).
This result is not expected to depend on the details of the model, such as the lattice and the properties of the initial state of the network, provided that the initial state is well connected. The only important features are the randomness and the quasistaticity of the breaking events.

We evolve the collapsed clusters back in time and analyze their initial configurations, similarly to the experiment (see also Ref. \cite{alvarado2013myosin}). The plot of initial configurations of all the clusters, presented in Fig. \ref{InitialAreas}(c), is similar to the cluster structure of a random percolation model \cite{stauffer1994introduction} at the critical point (see Fig. \ref{InitialAreas}(a)). The initial structure of the collapsed clusters, possessing scale-free enclaves, is fractal with a fractal dimension of the percolation model, $ d_\mathrm{f}'=d/ \left( \tau'-1 \right) = 91/48<2 $. We conclude that the \textit{phantom} model of the network collapse drives itself to the critical point of the \textit{random percolation model}. 

We turn now to consider effects of steric interactions and how they change the final state of the collapse from the critical point of the random percolation model to the one of the NEP model. Adding steric interactions to the phantom model, analyzed above, we get qualitative and quantitative agreements of the collapse model and the experiment. 

\subsection{Steric interactions}
In $2$-d networks steric interactions lead to collapse of enclaves to the same point as their surrounding, as illustrated in Fig. \ref{EnclaveInclusion}. Thus, one can take steric interactions into account by collapsing the phantom model, expanding the resulting clusters back in time, identifying enclaves and absorbing them into their surrounding clusters. Doing this, the number of collapsed clusters decreases and their distribution in space contains large voids (Fig. \ref{FinalState}(a,b)). Similar voids appear in the collapsed state of the experimental system, as shown in Fig. \ref{FinalState}(c). Taking steric interactions into account, the initial configuration of the collapsed clusters has no enclaves, as shown in Fig. \ref{InitialAreas}(d). The configuration is similar to the critical point of the NEP model's critical configuration, shown in Fig. \ref{InitialAreas}(b) and the initial configuration of the collapsed clusters in the experimental system, shown in Fig. \ref{Jose}(c). Moreover, the clusters' masses, in the case of collapse with steric interactions, are distributed as a power-law. The Fisher exponent is different from the one in the random percolation model, but indistinguishable from Eq. \eqref{tau}---the Fisher exponent of the NEP model (see Fig. \ref{Distributions}(b)) and the Fisher exponent obtained experimentally (see Fig. \ref{Jose}(d)). We conclude that our model for the collapse \textit{with steric interactions} drives itself into the transition point of the \textit{NEP model}. This is analogous to how the phantom model of collapse, ignoring steric interactions, drives itself into the critical point of a random percolation model.

\section{Discussion and summary}
This model of a collapse of active elastic networks with steric interactions reproduces two intriguing properties of the experimental system. Firstly, since the properties of the final state do not depend on the initial connectivity (provided it is well above percolation transition), it reproduces the robustness of the final state's criticality for a wide range of parameters. Secondly, it explains the qualitative disagreement of the criticality properties of the experiment with the ones in the random percolation model. In the experiment the enclaves in the initial configurations of the collapsed clusters are rare (see Fig. \ref{Jose}(c)) and the measured Fisher exponent is consistent with the NEP model's Fisher exponent.

Suppressing enclaves leads to an euclidean structure of the clusters with $\tau<2$ power-law distribution. In this paper we study the simplest mechanism of enclaves supression---steric interactions. However, the suppression of enclaves can be achieved by other mechanisms. Detailed modelling of the experimental system in Ref. \cite{alvarado2013myosin} also led to rare enclaves and a distribution of the collapsed clusters' 
masses with a power-law consistent with Eq. \eqref{tau}, although steric interactions were not taken into account. There the reduction of the local connectivity was coupled to the local stress, in contrast to random breaks used in this study. Since one expects a relatively small stress in a region where it is close to become an enclave (the stress on the last bond that connects the potential enclave to its potential surrounding cluster can be quickly relaxed), the generation of enclaves is suppressed by this coupling. Apparently, this suppression is sufficient to bring the system to the critical point of the NEP model. Analysis of the properties of the clusters' structure is not sufficient to distinguish between different mechanisms of enclaves suppression. However, the fact that enclaves are suppressed is quite evident from the experimental results.

Our model of the collapse with steric interactions is consistent with experiment in the regime where the final state is scale-free. However, other experimental regimes cannot be described by the model. The regime of small, collapsed clusters with exponential distribution, observed in Ref. \cite{alvarado2013myosin}, is not captured by the model due to its quasistaticity assumption. The global crack, observed in Ref. \cite{alvarado2013myosin}, is not captured by the model due to its assumption of the randomness of bond breaks. A more detailed model, without these assumptions, presented in Ref. \cite{alvarado2013myosin}, is able to describe all the observed regimes in the experiment. However, in contrast to that model, our minimalistic approach explains the qualitative differences of the collapsed state from the critical point of the random percolation model. Moreover, our approach highlights how steric interactions in active biological networks can affect physical properties of the system.

In this article we present a minimalistic model for collapse of an active network. The model drives itself to a special state. Ignoring steric interactions of the network elements, this state corresponds to the critical point of a random percolation model. A more realistic model of the network with steric interactions leads to a collapse to a state, which corresponds to a modified version of the percolation model---percolation model with enclaves absorbed in their surroundings (NEP model). The NEP model exhibits a Fisher exponent smaller than $ 2 $, at the transition point. The structural properties of the NEP model at the transition point are similar to experimentally observed cluster structures. Our model of collapse with steric interactions describes how a system robustly drives itself towards this point.
\acknowledgments
This work is part of the research programme of the Foundation for Fundamental Research on Matter (FOM), which is part of the Netherlands Organisation for Scientific Research (NWO). GK and JA were funded by a Vidi grant from the Netherlands Organization for Scientific
Research (NWO). The authors thank T. Michels for helpful discussions.

\bibliographystyle{apsrev}
\bibliography{SOCbib}

\end{document}